\documentclass[twocolumn,superscriptaddress,prl,floatfix,showpacs]{revtex4}
\usepackage{graphicx,bm}
\usepackage[psamsfonts]{amssymb}
\usepackage{type1cm}
\begin{document}

\title{Neutrino-induced nucleosynthesis of $A>64$ nuclei: The $\nu
  p$-process}

\author{C. Fr\"ohlich}
\affiliation{Departement f\"ur Physik und Astronomie, Universit\"at
  Basel, CH-4056 Basel, Switzerland}
\author{G. Mart\'{\i}nez-Pinedo}
\affiliation{ICREA and Institut d'Estudis Espacials de Catalunya,
  Universitat Aut\`onoma de Barcelona, E-08193 Bellaterra, Spain}
\affiliation{Gesellschaft f\"ur Schwerionenforschung, D-64291
  Darmstadt, Germany}
\author{M. Liebend\"orfer}
\affiliation{Canadian Institute for Theoretical Astrophysics, Toronto,
  ON M5S 3H8, Canada}
\affiliation{Departement f\"ur Physik und Astronomie, Universit\"at
  Basel, CH-4056 Basel, Switzerland}
\author{F.-K. Thielemann}
\affiliation{Departement f\"ur Physik und Astronomie, Universit\"at
  Basel, CH-4056 Basel, Switzerland}
\author{E. Bravo}
\affiliation{Departament de F{\'\i}sica i Enginyeria Nuclear,
  Universitat Polit\`ecnica de Catalunya, E-08034 Barcelona, Spain}
\author{W.~R.~Hix}
\affiliation{Physics Division, Oak Ridge National Laboratory, Oak
  Ridge, TN 37831}
\author{K. Langanke}
\affiliation{Gesellschaft f\"ur Schwerionenforschung, D-64291
  Darmstadt, Germany} 
\author{N. T. Zinner}
\affiliation{Institute for Physics and Astronomy, University of
  {\AA}rhus, DK-8000 {\AA}rhus C, Denmark}

\date{\today}

\begin{abstract}
  We present a new nucleosynthesis process, that we denote $\nu
  p$-process, which occurs in supernovae (and possibly gamma-ray
  bursts) when strong neutrino fluxes create proton-rich ejecta. In
  this process, antineutrino absorptions in the proton-rich
  environment produce neutrons that are immediately captured by
  neutron-deficient nuclei.  This allows for the nucleosynthesis of
  nuclei with mass numbers $A>64$. Making this process a possible
  candidate to explain the origin of the solar abundances of
  $^{92,94}$Mo and $^{96,98}$Ru.  This process also offers a natural
  explanation for the large abundance of Sr seen in an
  hyper-metal-poor star.
\end{abstract}

\maketitle

Supernova explosions (the cataclysmic endpoint of stellar evolution)
produce iron and neighboring nuclei, underlined by their lightcurves
powered by radioactive decay of $^{56}$Ni. The production of elements
beyond Fe has long been postulated by three classical processes, the
r- and the s-process (caused by rapid or slow neutron capture) and the
p-process, standing either for proton capture or alternative means to
produce heavy neutron deficient, stable
isotopes~\cite{Burbidge.Burbidge.ea:1957,Wallerstein.Iben.ea:1997}.
The s-process acts during stellar evolution via neutron captures on Fe
produced in previous stellar generations (thus being a ``secondary
process'').  The location and/or \emph{operation and uniqueness} of
the r- and p-process in astrophysical sites are still a subject of
debate.  The r-process is required to be a primary process in stellar
explosions~\cite{Sneden.Cowan:2003}, meaning that the production of
such elements is independent of the initial heavy element content in
the star.  Recent galactic chemical evolution studies of Sr, Y, and
Zr~\cite{Travaglio.Gallino.ea:2004} suggest the existence of a
\emph{primary process}, denoted ``lighter element primary process''
(LEPP), that is independent of the
r-process~\cite{Sneden.Cowan:2003,Cowan.Sneden.ea:2005} and operates
very early in the Galaxy. Most of the p-nuclei are thought to be
produced in hot (supernova) environments, where disintegration of
pre-existing heavy elements (thus being also a secondary process) due
to black-body radiation photons can account for the heavy $p$-nuclei
but underproduces the light ones (see e.g.\
ref.~\cite{Arnould.Goriely:2003,%
  Costa.Rayet.ea:2000,Hayakawa.Iwamoto.ea:2004}). Currently, the
mechanism for the production of the light $p$-nuclei, $^{92,94}$Mo and
$^{96,98}$Ru, is unknown, however chemical evolution studies of the
cosmochronometer nucleus $^{92}$Nb~\cite{Dauphas.Rauscher.ea:2003},
imply a primary supernova origin for these light $p$-nuclei.

Observations of extremely ``metal-poor'' stars in the Milky Way
provide us with information about the nucleosynthesis processes
operating at the earliest times in the evolution of our Galaxy. They
are thus probing supernova events from the earliest massive stars, the
fastest evolving stellar species. The recently discovered
hyper-metal-poor stars in the Milky
Way~\cite{Frebel.Aoki.ea:2005,Christlieb.Bessel.ea:2002} may witness
chemical enrichment by the first generation of faint massive
supernovae which experience extensive matter mixing (due to
instabilities) and fallback of matter after the
explosion~\cite{Iwamoto.Umeda.ea:2005}.  However, the detection of
Sr/Fe, exceeding 10 times the solar ratio, in the most metal-poor star
known to date~\cite{Frebel.Aoki.ea:2005} suggests the existence of a
primary process, producing elements beyond Fe and Zn.

In this Letter, we present a new nucleosynthesis process that will
occur in all core-collapse supernovae and could explain the existence
of Sr and other elements beyond Fe in the very early stage of galactic
evolution. We denote this process ``$\nu p$-process'' and suggest it
as a candidate for the postulated lighter element primary process
LEPP~\cite{Travaglio.Gallino.ea:2004}. It can also contribute to the
nucleosynthesis of light $p$-process nuclei. Here, we consider only
the inner ejecta of core-collapse supernovae, but the winds from the
accretion disk in the collapsar model of gamma-ray
bursts~\cite{Woosley:1993,Surman.Mclaughlin.Hix:2005,%
  Pruet.Thompson.Hoffman:2004,Fujimoto.Hashimoto.ea:2004} may also be
a relevant site for the $\nu p$-process.  The $\nu p$-process is
primary and is associated with explosive scenarios. It occurs when
strong neutrino fluxes create proton-rich ejecta. After the production
of Fe-group elements, continued antineutrino absorptions by free
protons produce free neutrons subject to immediate capture on
neutron-deficient nuclei with small proton-capture cross sections and
long beta-decay half-lives. This is distinct from earlier suggestions
for the production of $p$-nuclei in the so-called neutrino wind that
develops in later phases of a supernova
explosion~\cite{Fuller.Meyer:1995,Hoffman.Woosley.ea:1996}, because
the captured neutrons are directly delivered by antineutrino
absorption on free protons in a proton-rich environment.

As a full understanding of the core collapse supernova mechanism is
still pending and successful explosion simulations are difficult to
obtain~\cite{Buras.Rampp.ea:2005}, the composition of the innermost
ejecta -- directly linked to the explosion mechanism -- remained to a
large extent unexplored.  Recent supernova simulations with accurate
neutrino transport~\cite{Liebendoerfer.Mezzacappa.ea:2001a,%
Buras.Rampp.ea:2003,Thompson.Quataert.Burrows:2005} show the
presence of proton-rich neutrino-heated matter, both in the inner
ejecta~\cite{Liebendoerfer.Mezzacappa.ea:2001a,Buras.Rampp.ea:2003} and
in the early neutrino wind from the proto-neutron
star~\cite{Buras.Rampp.ea:2003}. This matter is subject to a large
neutrino energy deposition by the absorption of neutrinos and
antineutrinos with initially similar intensities and energy spectra.
As soon as the heating lifts the electron degeneracy, the reactions
$\nu_e + n \rightleftarrows p + e^{-}$ and $\bar{\nu}_e + p
\rightleftarrows n + e^{+}$ drive the composition proton-rich due to
the smaller mass of the proton~\cite{Froehlich.Hauser.ea:2005,%
Pruet.Woosley.ea:2005}, ($n$, $p$, $e^{-}$, $e^{+}$, $\nu_e$,
$\bar{\nu}_e$ denote the neutron, proton, electron, positron,
neutrino, and antineutrino respectively). Such proton-rich matter with
$Y_e$, the number of electrons or protons per nucleon, larger than 0.5
will always be present in core-collapse supernovae explosions with
ejected matter irradiated by a strong neutrino flux, independently of
the details of the explosion~\cite{Froehlich.Hauser.ea:2005}. As this
proton-rich matter expands and cools, nuclei can form resulting in a
composition dominated by $N=Z$ nuclei, mainly $^{56}$Ni and $^4$He,
and protons.  Without the further inclusion of neutrino and
antineutrino reactions the composition of this matter will finally
consist of protons, alpha-particles, and heavy (Fe-group) nuclei (in
nucleosynthesis terms a proton- and alpha-rich freeze-out), with
enhanced abundances of $^{45}$Sc, $^{49}$Ti, and
$^{64}$Zn~\cite{Froehlich.Hauser.ea:2005,Pruet.Woosley.ea:2005}.  In
these calculations the matter flow stops at $^{64}$Ge with a small
proton capture probability and a beta-decay half-life (64~s) that is
much longer than the expansion time scale ($\sim
10$~s)~\cite{Pruet.Woosley.ea:2005}. 

\begin{figure}
  \centering
  \includegraphics[width=\linewidth]{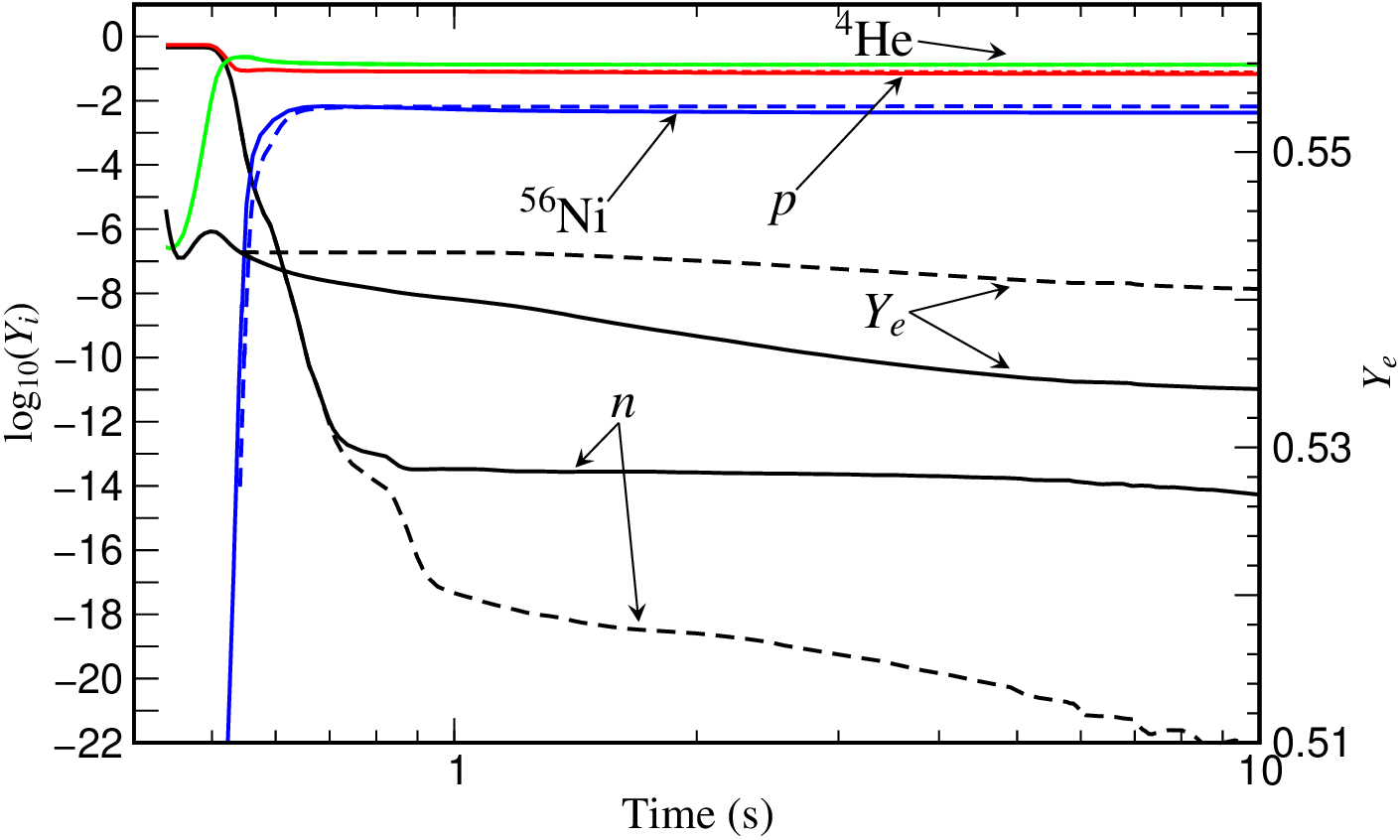}
  \caption{Evolution of the abundance of neutrons, protons,
    alpha-particles, and $^{56}$Ni in a nucleosynthesis trajectory
    resulting from model B07 of
    reference~\cite{Froehlich.Hauser.ea:2005}. The abundance, $Y$, is
    defined as the number of nuclei of the species $i$ present divided
    by the total number of nucleons which is conserved during the
    calculation. The solid (dashed) lines display the nucleosynthesis
    results which include (omit) neutrino and antineutrino absorption
    interactions after nuclei are formed. The abscissa measures the
    time since the onset of the supernova
    explosion.\label{fig:pnhe4evol}}
\end{figure}

Synthesis of nuclei heavier than $A=64$ is possible in proton rich
ejecta if the entropy per nucleon is in the range $s \approx
150$--170~$k_B$ (where $k_B$ is the Boltzmann
constant)~\cite{Jordan.Meyer:2004}.  Such large entropies are,
however, not attained in core-collapse supernovae simulations with
detailed neutrino transport which give $s\approx
50$--75~$k_B$~\cite{Froehlich.Hauser.ea:2005,Pruet.Woosley.ea:2005}.
Here we show that the synthesis of nuclei with $A>64$ can also be
obtained with realistic entropies, if one explores the previously
neglected effect of neutrino interactions on the nucleosynthesis of
heavy nuclei. When interactions with neutrinos and antineutrinos are
considered for both free and bound nucleons the situation becomes
dramatically different. $N \sim Z$ nuclei are practically inert to
neutrino capture (converting a neutron in a proton), because such
reactions are endoergic for neutron-deficient nuclei located away from
the valley of stability.  The situation is different for antineutrinos
that are captured in a typical time of a few seconds, both on protons
and nuclei, at the distances at which nuclei form ($\sim 1000$~km).
This time is much shorter than the beta-decay half-life of the most
abundant heavy nuclei reached without neutrino interactions (e.g.\
$^{56}$Ni, $^{64}$Ge). As protons are more abundant than heavy nuclei,
antineutrino capture occurs predominantly on protons, causing a
residual density of free neutrons of $10^{14}$--$10^{15}$~cm$^{-3}$
for several seconds, when the temperatures are in the range 1--3~GK\@.
This effect is clearly seen in figure~\ref{fig:pnhe4evol}, where the
time evolution of the abundances of protons, neutrons, alpha-particles
and $^{56}$Ni is shown ($^{56}$Ni serves to illustrate when nuclei are
formed). The dashed lines shows the results for a calculation where
neutrino absorptions are neglected once the temperature drops below
6~GK. This allows to study the effect of neutrino absorptions in the
latter phase of nucleosynthesis when the $\nu p$-process acts without
changing the initial phases where $Y_e$ is determined. Without the
inclusion of antineutrino capture the neutron abundance soon becomes
too small to allow for any capture on heavy nuclei. The figure
also compares the evolution of $Y_e$.

In our studies we use the detailed neutrino spectral information
provided by neutrino radiation hydrodynamical calculations to
determine the neutrino antineutrino absoption rates at each point of
the nucleosynthesis trajectory (temperature, density and radius).  Our
network calculations follow the detailed abundances of 1435 isotopes
between $Z=1$ and $Z=54$, which allows an accurate treatment of the
changes in composition induced by neutrino interactions. However, our
network calculations follow the $Y_e$ evolution of the hydrodynamical
calculations only till the moment when alpha particles form.  At this
time, the determination of the $Y_e$ value in the hydrodynamical
studies is plagued by an error in the Lattimer-Swesty equation of
state~\cite{Lattimer.Swesty:1991,Lattimer:2004}, which we had adopted
in the hydrodynamical calculations. This error results in an
underproduction of alpha particles which suppresses the occurence of
an alpha-effect which drives the $Y_e$ value closer to
0.5~\cite{Fuller.Meyer:1995,Meyer.Mclaughlin.Fuller:1998}.  While such
alpha effect does not occur in the hydrodynamical calculations as the
computed alpha abundance is too low, it is present in the network
calculations.  However, we stress that, in contrast to
expectation~\cite{Pruet.Woosley.ea:2005}, the alpha effect is no
obstacle for the nucleosynthesis of heavy nuclei, because once heavy
nuclei form the neutrons are captured by heavy neutron-deficient
nuclei instead of forming deuterium and later alpha particles.

The neutrons produced via antineutrino absorption on protons can
easily be captured by neutron-deficient $N\sim Z$ nuclei (for example
$^{64}$Ge), which have large neutron capture cross sections. The amount
of nuclei with $A>64$ produced is then directly proportional to the
number of antineutrinos captured.  While proton capture, $(p,\gamma)$,
on $^{64}$Ge takes too long, the $(n,p)$ reaction dominates (with a
lifetime of 0.25~s at a temperature of 2~GK), permitting the matter
flow to continue to nuclei heavier than $^{64}$Ge via subsequent
proton captures with freeze-out at temperatures around 1~GK\@. This is
different to r-process environments with $Y_e < 0.5$, i.e.\
neutron-rich ejecta, where neutrino capture on neutrons provides
protons that interact mainly with the existing neutrons, producing
alpha-particles and light nuclei. Their capture by heavy nuclei is
suppressed because of the large Coulomb
barriers~\cite{Fuller.Meyer:1995,Meyer.Mclaughlin.Fuller:1998}.
Consequently, in r-process studies an enhanced formation of the
heaviest nuclei does not take place when neutrino interactions are
included. In proton-rich ejecta, however, antineutrino absorption
produces neutrons that do not suffer from Coulomb barriers and are
captured preferentially by heavy neutron-deficient nuclei.

\begin{figure}
  \centering
  \includegraphics[width=\linewidth]{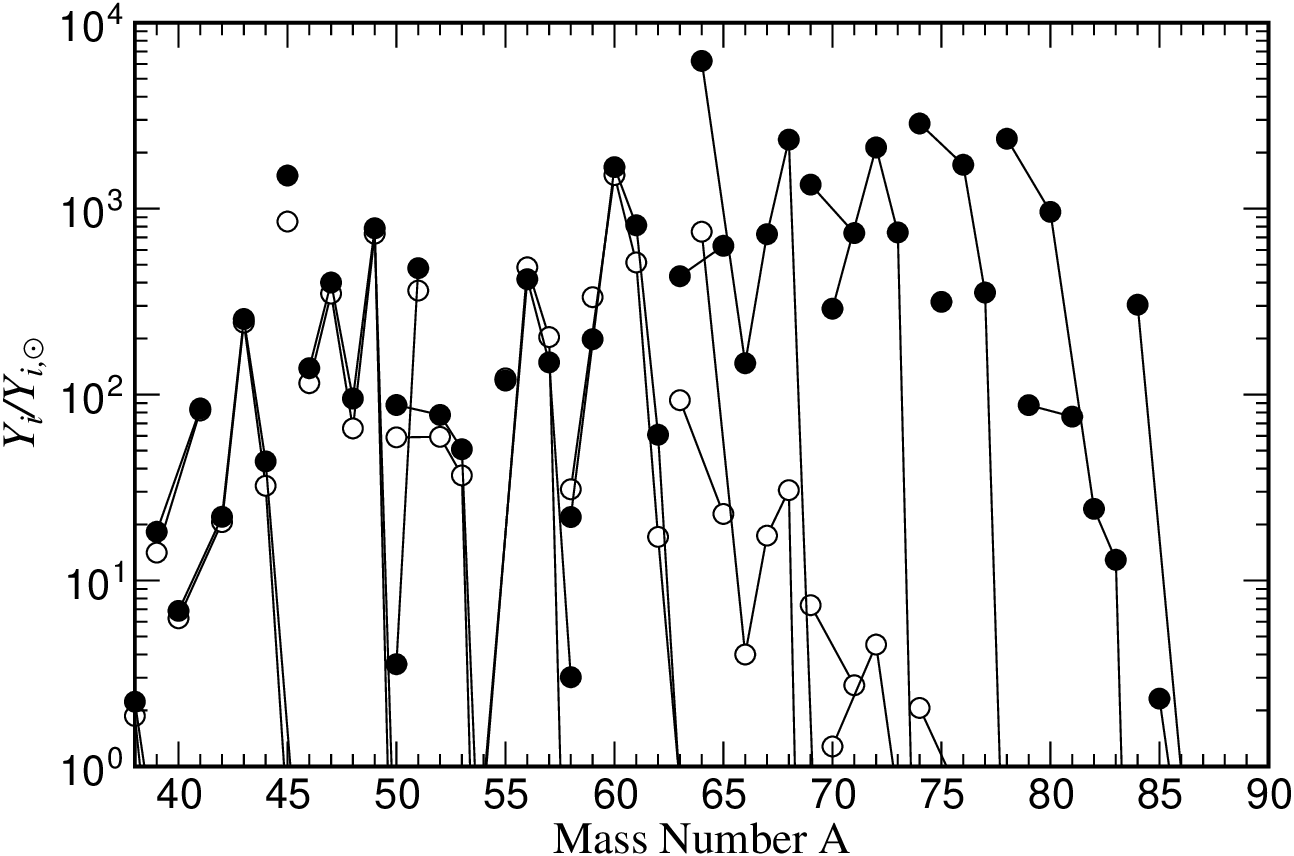}
  \caption{The panel shows the isotopic abundances for model B07 of
    reference~\cite{Froehlich.Hauser.ea:2005} relative to solar
    abundances~\cite{Lodders:2003}. The filled circles represent
    calculations where (anti)neutrino absorption reactions are
    included in the nucleosynthesis while for the open circles
    neutrino interactions are neglected.\label{fig:abund}}
\end{figure}

Figure~\ref{fig:abund} shows the composition of supernova ejecta
obtained with the hydrodynamical model B07 described in detail in
ref.~\cite{Froehlich.Hauser.ea:2005}. In addition to the proton-rich
conditions in the innermost ejected zones visible in simulations by
different groups~\cite{Liebendoerfer.Mezzacappa.ea:2001a,%
  Buras.Rampp.ea:2003,Thompson.Quataert.Burrows:2005}, our models
consistently include neutrino-absorption reactions in the
nucleosynthesis calculations allowing for the occurrence of the $\nu
p$-process. However, in our stratified spherically symmetric models
the accretion rate is rapidly reduced (and with this the neutrino
luminosities) with the onset of the explosion. In a more realistic
scenario considering convective turnover in the hot mantle, continued
accretion is expected to maintain a large neutrino luminosity beyond
the onset of the explosion and to further support the $\nu p$-process

In order to understand the sensitivity of our results one must
consider the dependence of the $\nu p$-process on the conditions
during the ejection of matter in supernova explosions. There are
several essential parameters in addition to the entropy $s$. One is
the $Y_e$-value of the matter when nuclei are formed. The larger the
$Y_e$-value, the larger is the proton abundance, producing a larger
neutron abundance for the same antineutrino flux during the $\nu
p$-process.  This permits a more efficient bridging of beta-decay
waiting points by $(n,p)$-reactions in the flow of proton captures to
heavier nuclei. The location (radius $r$) of matter during the
formation of nuclei and the ejection velocity also influence the $\nu
p$-process by determining the intensity and duration of the
antineutrino flux which the matter will experience. A location closer
to the surface of the protoneutron star and/or a slow ejection
velocity leads to an extended antineutrino exposure that allows for an
increased production of heavy elements.  Finally, the long-term
evolution of the neutrino luminosities and energy spectra during the
cooling phase of the proto-neutron star plays an important role. These
factors are poorly known due to existing uncertainties in the
supernova explosion mechanism.

\begin{figure}
  \centering
  \includegraphics[width=\linewidth]{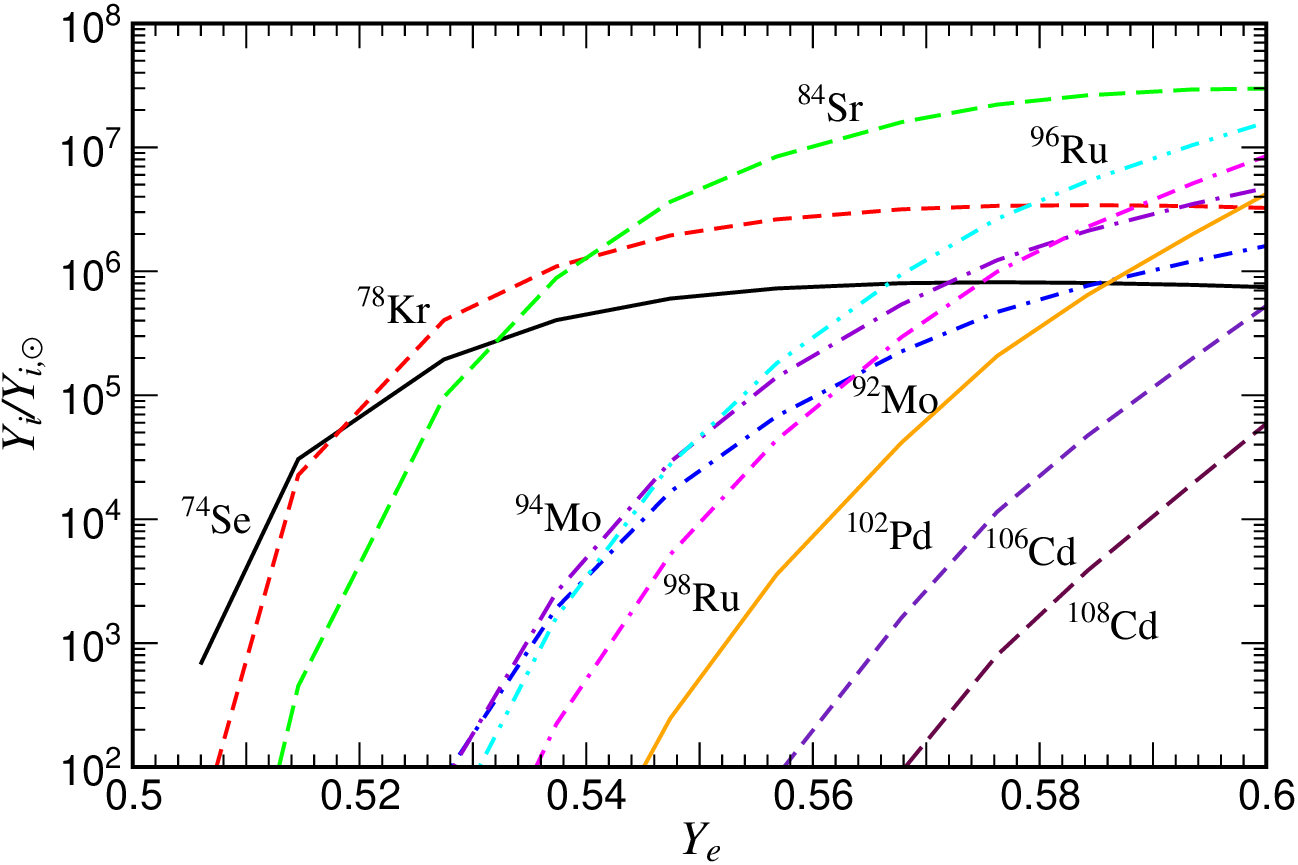}
  \caption{Light $p$-nuclei abundances in comparison to solar
    abundances as a function of $Y_e$. The $Y_e$-values given are the
    ones obtained at a temperature of 3~GK that corresponds to the
    moment when nuclei are just formed and the $\nu p$-process starts
    to act.\label{fig:ye}}
\end{figure}

To test the dependence of the nucleosynthesis on these parameters we
have also carried out parametric calculations based on adiabatic
expansions similar to those used in
refs.~\cite{Jordan.Meyer:2004,Meyer:2002} but for a constant realistic
entropy per nucleon $s=50\ k_B$.  This allows exploration of the
sensitivity of the nucleosynthesis without the need to perform full
radiation-hydrodynamical calculations. An example is given in
figure~\ref{fig:ye} which shows the dependence of the $p$-nuclei
abundances as a function of the $Y_e$ value of the ejected matter. The
different $Y_e$ values have been obtained by varying the temperatures
of the neutrino and antineutrino spectra assuming Fermi-Dirac
distributions for both. Close to $Y_e=0.5$ (and below) essentially no
nuclei beyond $A=64$ are produced. Nuclei heavier than $A=64$ are only
produced for $Y_e>0.5$, showing a very strong dependence on $Y_e$ in
the range 0.5--0.6.  A clear increase in the production of the light
$p$-nuclei, $^{92,94}$Mo and $^{96,98}$Ru, is observed as $Y_e$ gets
larger. However, additional work is necessary, including an
exploration of the nuclear physics uncertainties, to decide whether or
not the solar system abundances of light p-nuclei are due to the $\nu
p$-process. We obtain similar variations (not shown in the figure) in
the production of Sr, Y and Zr, in agreement with the large scatter of
the abundances seen in low metallicity
stars~\cite{Travaglio.Gallino.ea:2004}.

After we have presented our results in several conferences, the $\nu
p$-process has also been explored by Pruet \emph{et
  al}~\cite{Pruet.Hoffman.ea:2005}, who confirm our findings. They
have considered the nucleosynthesis in the early proton-rich neutrino
wind in a supernova explosion. The conditions there, larger entropies
($\sim 75$~kb) and nuclei forming at shorter distances (several
hundreds km) from the neutron star, extend the process described here
to even heavier nuclei.

All core-collapse supernova explosions, independent of existing model
uncertainties, will eject hot, explosively processed matter subject to
strong neutrino irradiation. We argue that in all cases the $\nu
p$-process will operate in the innermost ejected layers producing
neutron deficient nuclei above $A=64$. As the innermost ejecta, this
matter is most sensitive to the details of individual explosions, thus
their abundances will vary noticeable from supernova to supernova
(e.g.\ as a function of stellar mass, rotation, etc.). The final
amount of matter ejected will also depend on the intensity of the
fallback, but as discussed in ref.~\cite{Iwamoto.Umeda.ea:2005},
mixing before fallback will always lead to the ejection of elements
synthesized, even in the innermost layers.
Ref.~\cite{Iwamoto.Umeda.ea:2005} explains the abundances seen in HMP
stars by the ejecta of faint/weak core-collapse supernovae.  Such
faint supernova will generally have small expansion velocities
favoring an enhanced production of $\nu p$-elements, offering an
explanation for the presence of Sr in the star HE
1327-2326~\cite{Frebel.Aoki.ea:2005}. Our studies of the $\nu
p$-process show that the elements between Zn and Sr should be
co-produced together with Sr. The observation of these elements, which
with the exception of Ge and Rb are not detectable from the ground in
optical lines, but possible from space in the infrared or near
ultraviolet (e.g.\ the Hubble Space and Spitzer Space telescopes), can
provide support for the occurrence of the $\nu p$-process at early
times in the Galaxy. Such observations can contribute valuable
information about the conditions experienced by the inner supernova
ejecta in order to constrain current theoretical models of supernova
explosions.  Further studies are required to fully understand the $\nu
p$-process contribution to the chemical evolution of the Galaxy. In
this paper, we have discussed the innermost proton-rich supernova
ejecta before the emergence of the neutrino-wind from the protoneutron
star. This neutrino-wind will initially be
proton-rich~~\cite{Pruet.Woosley.ea:2005} but will turn neutron-rich
in its later phases allowing for the synthesis of r-process
nuclei~~\cite{Meyer.Mclaughlin.Fuller:1998,Hoffman.Woosley.ea:1996,%
  Fuller.Meyer:1995}. The variations in the contribution of the $\nu
p$-process (represented by Sr, Y, and Zr) and the r-process (producing
the heaviest elements up to Th and U)~\cite{Sneden.Cowan:2003,%
  Cowan.Sneden.ea:2005} can shed light on the connection of both of
these processes and provide information about the class of supernovae
that produce the heavy r-process nuclei.

\begin{acknowledgments}
  CF and FKT are supported by the Swiss National Science Foundation
  (SNF). GMP is supported by the Spanish MEC and by the European Union
  ERDF under contracts AYA2002-04094-C03-02 and AYA2003-06128. ML is
  suported by the SNF under grant PP002--106627/1. EB research has
  been supported by DURSI of the Generalitat de Catalunya and Spanish
  DGICYT grants. WRH research has been supported in part by the United
  States National Science Foundation under contract PHY-0244783 and by
  the United States Department of Energy, through the Scientific
  Discovery through Advanced Computing Program. Oak Ridge National
  Laboratory is managed by UT-Battelle, LLC, for the U.S. Department
  of Energy under contract DE-AC05-00OR22725.
\end{acknowledgments}

\bibliography{bibliography}

\end{document}